\documentclass[11pt,twocolumn,prb,superscriptaddress,reprint]{revtex4-1}

\usepackage{graphicx,amsmath,amssymb,color}
\usepackage{tabularx, ctable}
\usepackage{ulem}
\usepackage{bm}
\usepackage{siunitx}
\usepackage{gensymb}
\usepackage{wrapfig}
\usepackage{dirtytalk}
\usepackage{xcolor}

\usepackage[breaklinks=true,colorlinks,allcolors=blue]{hyperref}

\newcommand{\be}{\begin{eqnarray}}
\newcommand{\ee}{\end{eqnarray}}

\begin{document}

\title{Anomalous terahertz photoconductivity caused by the \\ superballistic flow of hydrodynamic electrons in graphene}


\author{M. Kravtsov}
\affiliation{Institute for Functional Intelligent
Materials, National University of Singapore, Singapore, 117575, Singapore}

\author{A. L. Shilov}
\affiliation{Institute for Functional Intelligent
Materials, National University of Singapore, Singapore, 117575, Singapore}
\affiliation{Programmable Functional Materials Lab, Center for Neurophysics and Neuromorphic Technologies, Moscow, 127495}

\author{Y. Yang}
\affiliation{Institute for Functional Intelligent
Materials, National University of Singapore, Singapore, 117575, Singapore}

\author{T. Pryadilin}
\affiliation{Department of Materials Science and Engineering, National University of Singapore, 117575 Singapore}

\author{M. A. Kashchenko}
\affiliation{Programmable Functional Materials Lab, Center for Neurophysics and Neuromorphic Technologies, Moscow, 127495}
\affiliation{Center for Photonics and 2D Materials, Moscow Institute of Physics and Technology, Dolgoprudny, 141700}

\author{O. Popova}
\affiliation{Programmable Functional Materials Lab, Center for Neurophysics and Neuromorphic Technologies, Moscow, 127495}

\author{M. Titova}
\affiliation{Programmable Functional Materials Lab, Center for Neurophysics and Neuromorphic Technologies, Moscow, 127495}

\author{D. Voropaev}
\affiliation{Programmable Functional Materials Lab, Center for Neurophysics and Neuromorphic Technologies, Moscow, 127495}

\author{Y. Wang}
\affiliation{Institute for Functional Intelligent
Materials, National University of Singapore, Singapore, 117575, Singapore}

\author{K. Shein}
\affiliation{Moscow Pedagogical State University, Moscow 119991}
\affiliation{National Research University Higher School of Economics, Moscow, 101000.}

\author{I. Gayduchenko}
\affiliation{Moscow Pedagogical State University, Moscow 119991}
\affiliation{National Research University Higher School of Economics, Moscow, 101000.}

\author{G.N. Goltsman}
\affiliation{Moscow Pedagogical State University, Moscow 119991}
\affiliation{National Research University Higher School of Economics, Moscow, 101000.}

\author{M. Lukianov}
\affiliation{Programmable Functional Materials Lab, Center for Neurophysics and Neuromorphic Technologies, Moscow, 127495}

\author{A. Kudriashov}
\affiliation{Institute for Functional Intelligent
Materials, National University of Singapore, Singapore, 117575, Singapore}
\affiliation{Programmable Functional Materials Lab, Center for Neurophysics and Neuromorphic Technologies, Moscow, 127495}

\author{T.~Taniguchi}
\affiliation{International Center for Materials Nanoarchitectonics, National Institute of Material Science, Tsukuba 305-0044, Japan}

\author{K.~Watanabe}
\affiliation{Research Center for Functional Materials, National Institute of Material Science, Tsukuba 305-0044, Japan}

\author{D.A.~Svintsov}
\affiliation{Center for Photonics and 2D Materials, Moscow Institute of Physics and Technology, Dolgoprudny, 141700}

\author{A. Principi}
\affiliation{School of Physics and Astronomy, University of Manchester, Manchester M13 9PL, United Kingdom}

\author{S. Adam}
\affiliation{Department of Materials Science and Engineering, National University of Singapore, 117575 Singapore}

\author{K.S. Novoselov}
\affiliation{Institute for Functional Intelligent
Materials, National University of Singapore, Singapore, 117575, Singapore}

\author{D. A. Bandurin$^{*}$}
\affiliation{Department of Materials Science and Engineering, National University of Singapore, 117575 Singapore}


\begin{abstract}
Light incident upon materials can induce changes in their electrical conductivity, a phenomenon referred to as photoresistance. In semiconductors, the photoresistance is negative, as light-induced promotion of electrons across the band gap enhances the number of charge carriers participating in transport. In superconductors, the photoresistance is positive because of the destruction of the superconducting state, whereas in normal metals it is vanishing. Here we report a qualitative deviation from the standard behavior in metallic graphene. We show that Dirac electrons exposed to continuous wave (CW) terahertz (THz) radiation can be thermally decoupled from the lattice by 50~K which activates hydrodynamic electron transport. In this regime, the resistance of graphene constrictions experiences a decrease caused by the THz-driven superballistic flow of correlated electrons. We analyze the dependencies of the negative photoresistance on the carrier density, and the radiation power and show that our superballistic devices operate as sensitive phonon-cooled bolometers and can thus offer a picosecond-scale response time. Beyond their fundamental implications, our findings underscore the practicality of electron hydrodynamics in designing ultra-fast THz sensors and electron thermometers.
\end{abstract}

\maketitle

Photons absorbed by graphene can induce a significant increase in its electronic temperature, $T_{\text{e}}$, owing to the remarkably low specific heat capacity of Dirac electrons~\cite{Specific_heat_Efetov} and weak electron-phonon interaction~\cite{KC_noise,Noise_cooling,eph_coupling,eph_coupling2}. These properties, coupled with rapid quasiparticle thermalization~\cite{hot_carriers}, have promised graphene-based ultra-fast light sensors and broadband cameras~\cite{koppens_photodetectors_2014,bonaccorso_graphene_2010}. However, the immunity of doped graphene's resistance to variations in $T_{\text{e}}$ has posed difficulties in utilizing these properties for sensitive bolometers, particularly in the technically demanding THz range~\cite{2D_THZ_REVIEW}. Indeed, since $T_{\text{e}}$-sensitive electron-electron (e-e) umklapp collisions are drastically suppressed in graphene~\cite{Wallbank2019}, the rise of $T_{\text{e}}$ relative to the lattice temperature does not lead to changes in its conductivity, which is only limited by electron-phonon scattering~\cite{Efetov_eph,Rosh_MP,Wang2013c}. This necessitates the use of indirect $T_{\text{e}}$-to-voltage conversion schemes such as Josephson junctions~\cite{Lee2020}, thermoelectric p-n junctions~\cite{Bandurin_dual,Castilla2019,Titova2023}, Johnson noise readout~\cite{Efetov2018}, and quantum dots~\cite{QD_THz}, to harness unique properties of graphene in sensing electromagnetic radiation.

\begin{figure*}[ht!]
  \centering\includegraphics[width=0.9\linewidth]{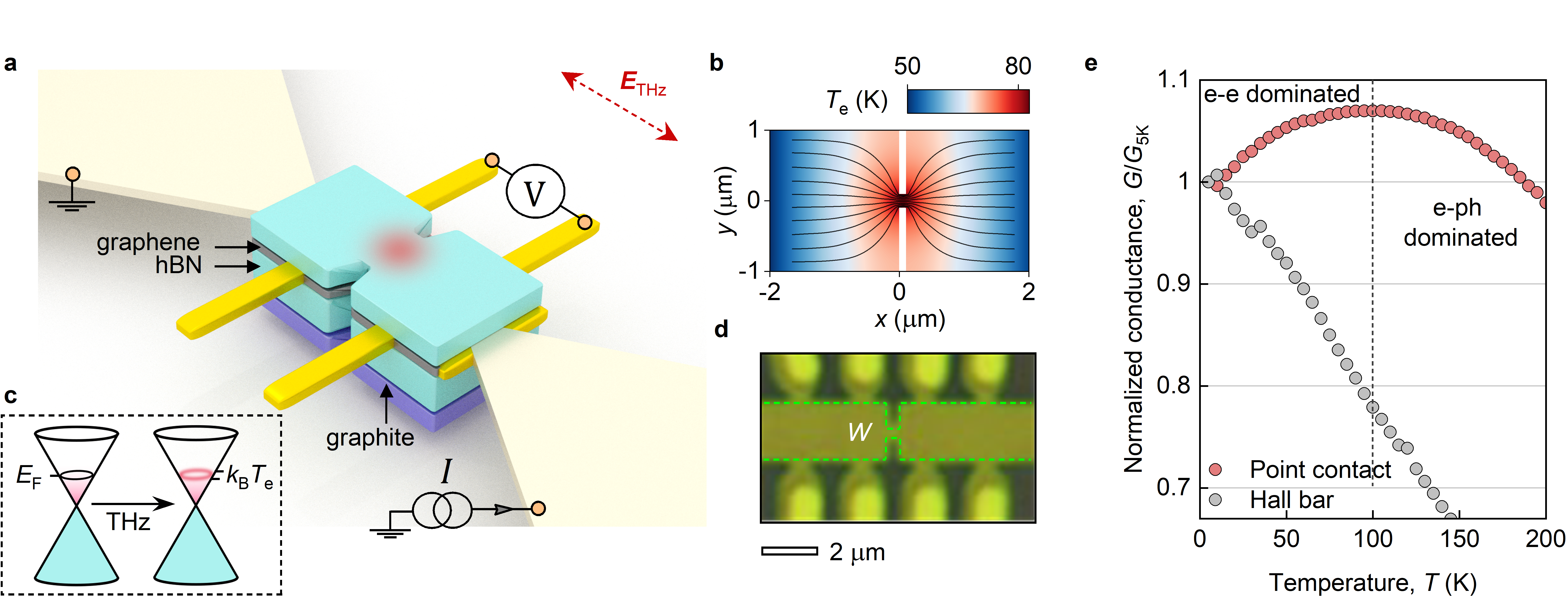}
    \caption{\textbf{Superballistic electron flow.} \textbf{a,} Schematic illustration of the device architecture: graphene PC is coupled to a broadband bow-tie antenna exposed to THz radiation. Absorbed radiation causes the increase of $T_\mathrm{e}$ while leaving the lattice $T$ intact. At the center of the device, $T_\mathrm{e}$ is higher than at the sample boundaries where it is thermalized with the bath. \textbf{b,} Electron temperature, $T_\mathrm{e}$, mapped onto the streamlines of electrical current flowing through the PC exposed to THz radiation.  \textbf{c,} Schematic illustrating THz-induced heating of electrons in doped graphene. \textbf{d,} Photograph of one of our PC devices. \textbf{e,} Conductance as a function of $T$ for the PC and the Hall bar measured in the dark at given $n$. The non-monotonic $T-$dependence found for the PC signals e-e dominated superballistic conduction of hydrodynamic electrons~\cite{levitov2016,kumar2017superball}. }
    \label{fig:F1}
\end{figure*}

Hydrodynamic electron transport presents a natural yet hitherto overlooked alternative to this inquiry. This transport regime has been predicted to emerge in materials in which electron-electron (\textit{e-e}) collisions are momentum-conserving and their rate is the fastest among other scattering processes in the system~\cite{gurzhi1968,de_jong_hydrodynamic_1995}. Recent experiments on graphene materialized this regime and revealed the manifestation of electron hydrodynamics in the transport properties of graphene-based heterostructures~\cite{Lucas,Hydro-review}. Notable hydrodynamic effects predicted and observed in graphene include the violation of the Wiedeman-Franz law~\cite{Crossno}, negative non-local resistance~\cite{bandurin2016,bandurin2018}, Hall viscosity~\cite{berdyugin2019hydro}, Poiselle flow~\cite{Sulpizio2019,Yacobi_hydro}, quantum-critical conductivity~\cite{QuantumCritFritz,Schmalian,Gallagher}, electron-hole 
 friction~\cite{Nam2017,Shaffique_eh,Bandurin_PRL-TBG2}, giant thermal diffusivity~\cite{heatspread}, and hydrodynamic plasmons~\cite{Zhao2023,Koppens_hydro-plasmons}, to name a few. Central to all these effects is their sensitivity to $T_\mathrm{e}$ through the rate of momentum-conserving e-e collisions $1/\tau_\textit{e-e}\approx(k_\mathrm{B}T_\mathrm{e})^2/\hbar E_\mathrm{F}$, where $E_\mathrm{F}$ is the Fermi energy and $k_\mathrm{B}$ and $\hbar$ are the Boltzmann and Planck constants, respectively~\cite{BulkShear}. This observation renders hydrodynamic transport coefficients a direct measure of $T_\mathrm{e}$ and hence may enable the design of THz-sensitive devices. 

To gain some qualitative insight into the behavior of a hydrodynamic electron fluid subjected to incident THz radiation, we calculated $T_\mathrm{e}$ distribution map together with the current streamlines in a special measurement geometry that amplifies the effects of hydrodynamics -- a narrow graphene constriction also referred to as a (classical) point contact (PC), see Fig.~1a for schematic. The results are shown in Fig.~1b. Absorbed THz radiation results in the local elevation of $T_\mathrm{e}$ (Fig.~1c) above the cold lattice, especially in the constriction region where the THz-induced current density is highest (Fig.~1b). This, in turn, leads to the enhancement of the PC conductance caused by the superballistic electron transport in which the correlated flow of electrons prevents individual fermions from momentum relaxation at sample boundaries due to a streaming effect~\cite{Levitov_anti_Matthiessen,kumar2017superball}.


In this work, we observed this conduction boost in high-mobility graphene devices endowed with PCs and coupled to a broadband antenna exposed to THz radiation. A detailed analysis of the THz-driven superballistic conduction as a function of carrier density, PC width, and THz power reveals that our antenna-coupled PCs act as phonon-cooled electron bolometers and can thus offer picosecond-scale temporal response~\cite{hot_carriers}. The hydrodynamic bolometer was responsive to THz radiation up to 100~K - temperature at which momentum-relaxing electron-phonon (e-ph) scattering starts to suppress hydrodynamic electron transport and provide an efficient electron cooling path. Next, using the superballistic conduction model, we extracted $T_\mathrm{e}$ as a function of absorbed THz power and lattice temperature, $T$, which allowed us to measure the thermal conductance of hydrodynamic fluid across the PC. This methodology thus provides a convenient alternative to the Johnson thermometry approach used to investigate the thermal properties of graphene devices~\cite{KC_noise,Betz2013}. Beyond their fundamental implications, our findings underscore the practicality of electron hydrodynamics in designing ultra-fast THz sensors and electron thermometers.

\begin{figure*}[ht!]
  \centering\includegraphics[width=0.99\linewidth]{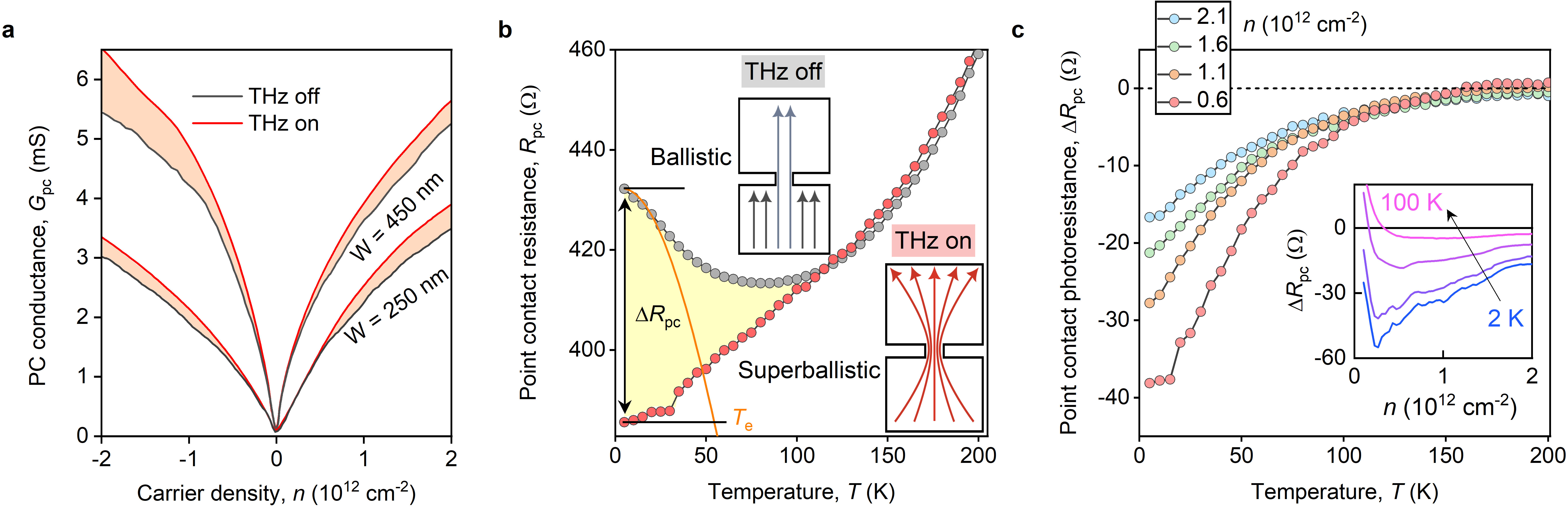}
    \caption{\textbf{THz-driven hydrodynamics and negative photoresistance.} \textbf{a,} PC conductance as a function of carrier density, $n$, measured in the dark and upon CW exposure with 0.14~THz radiation. The data was obtained for two PCs of a given width. $T=2~$K. \textbf{b,} $R_\mathrm{pc}$ as a function of $T$ in the dark (THz off) and upon exposing the device with 0.14 THz radiation (THz on). The yellow region between THz on and THz off region represents the negative photoresistance $\Delta R_\mathrm{pc}$. Instet: Schematical illustration of the ballistic non-interacting (upper inset) and superballistic collective electron motion (lower inset). $n=0.5\times10^{12}$ cm$^{-2}$. \textbf{c,} $\Delta R_\mathrm{pc}$ as a function of temperature $T$ for given carrier densities $n$ measured using the double-modulation technique (Methods). Inset: $\Delta R_\mathrm{pc}$ as a function of carrier density $n$ for different $T$.}
    \label{fig:F2}
\end{figure*}

\textbf{Design and characterization.} Our devices are monolayer graphene PCs coupled to a bow-tie antenna designed to operate in sub-THz and THz frequency domains (Fig.~1a). The devices were made using a standard dry transfer technique~\cite{CleaningInterfaces}. Through this method, graphene was encapsulated between two slabs of hexagonal boron-nitride (hBN) crystals and equipped with graphite bottom gates that both controlled the carrier density, $n$, in the channel and ensured low charge disorder. The device design was a  2~$\mu \mathrm{ m}$-wide graphene channel split into two sections by a sub-micron PC positioned between adjacent voltage probes; two constrictions with widths of $W=250$ and $450$~nm were studied (Fig. \ref{fig:F1}d). On each side of the constriction, additional voltage probes were patterned to independently characterize the properties of the graphene channel (Fig~\ref{fig:F1}d). 

Prior to photoresistance measurements, we characterize the transport properties of our devices. Both devices feature high-mobility, $\mu$, electron transport with $\mu$ reaching 10 m$^2$/Vs at $T=2~$K determined from the measurements of wide reference regions (Supplementary Information). Figure~\ref{fig:F1}e shows an example of the temperature ($T$) dependence of the PC conductance, $G_\mathrm{pc}$, measured at $n=0.5\times10^{12}$ cm$^\mathrm{-2}$. At $T=2\mathrm{~K}$, $G_\mathrm{pc}$ is close to the Sharvin point contact limit $G_0=4e^2W\sqrt{\pi n}/\pi h$ for the constriction $W=0.45~\mathrm{\mu m}$~\cite{sharvin1965possible,size-quant}. As $T$ rises, $G_\mathrm{pc}$ gradually increases, reaches maximal value, and starts to drop when $T$ is above 100 K. The initial increase of $G_\mathrm{pc}$ above $G_\mathrm{0}$ is the manifestation of the interaction-dominated electron transport in graphene PCs and is enabled by the superballistic conduction mechanism~\cite{Levitov_anti_Matthiessen,kumar2017superball}. 
The boost of the PC conduction above the single-particle Sharvin limit is described by the Levitov-Falkovich anti-Matthiessen rule~\cite{Levitov_anti_Matthiessen,kumar2017superball}: 
\begin{equation}\label{anti-M}
G_\mathrm{pc} = G_\mathrm{0}+G_\mathrm{\nu},
\end{equation}
where $G_\mathrm{\nu} =  W^2 e^2 v_\mathrm{F}  \sqrt{\pi |n|}/32 \hbar \nu$ is the viscous contribution to the PC conductance. Here $v_\mathrm{F}$ is the Fermi velocity and $\nu=v_\mathrm{F}^2\tau_\textit{e-e}/4$ is the kinematic viscosity of electron liquid controlled by $T_\mathrm{e}$. In transport experiments, $T_\mathrm{e}$ is varied upon changing the sample $T$ that, at elevated temperatures, also involves the enhancement of e-ph scattering, resulting in a non-monotonic $T-$dependence in Fig.~\ref{fig:F1}e. This dependence contrasts the behavior observed in a wide reference region whose conductance is insensitive to e-e collisions and which exhibits a standard monotonic drop of conductance with increasing $T$ due to e-ph scattering~\cite{Efetov_eph,Wang2013c} (Fig.~\ref{fig:F1}e).  

\textbf{Photoresponse measurements.} The central result of our study is presented in Fig.~\ref{fig:F2}a. It shows $G_\mathrm{pc}$ as a function $n$ measured in the dark and upon exposing the devices with 0.14 THz radiation; the samples were kept at $T=2~$K (see Methods for the measurements details). For both $W$, $G_\mathrm{pc}$ experiences a clear surge for both electron and hole doping when the PCs are exposed to THz radiation. The enhancement of conductance upon illumination, typical for semiconductors, is unexpected for metallic systems and can only be caused by the radiation-induced change of $T_\mathrm{e}$ that activates the hydrodynamic electron flow illustrated in the inset of Fig.~\ref{fig:F2}b. We note in passing, that antenna-coupled wide graphene channels, as expected, do not feature any photoresistance away from the charge neutrality point in agreement with previous studies~\cite{Mylnikov2023} (Supplementary Information).

Figure \ref{fig:F2}b compares the PC resistance, $R_\mathrm{pc}=1/G_\mathrm{pc}$, measured upon varying sample $T$ in the dark and when the THz excitation is applied. A clear difference between the two persists even above $100~$K, a characteristic temperature at which phonons start to destroy the hydrodynamic flow by momentum-relaxing collisions~\cite{bandurin2016}.
It is instructive to isolate the THz-sensitive part of the PC resistance. To this end, we employed a dual-modulation technique (Methods) and recorded the photoresistance $\Delta R_\mathrm{pc}=R_\mathrm{pc}^\mathrm{on}-R_\mathrm{pc}^\mathrm{off}$ as a function of $n$ and $T$ (here \textit{on} and \textit{off} denote signals recorded, respectively, in the dark and upon THz exposure). Figure~\ref{fig:F2}c shows that the superballistic photoresistance $\Delta R_\mathrm{pc}$ is negative for all $n$ and vanishes above $\approx120-130~K$. The inset of Fig.~\ref{fig:F2}c reveals that the dependence of $\Delta R_\mathrm{pc}$ on $n$ is predominantly non-monotonic: $\Delta R_\mathrm{pc}$ gradually decreases for $n>0.5\times10^{-12}$ whereas for smaller $n$ it grows rapidly and becomes positive at elevated $T$. 


\begin{figure*}[ht!]
  \centering\includegraphics[width=0.99\linewidth]{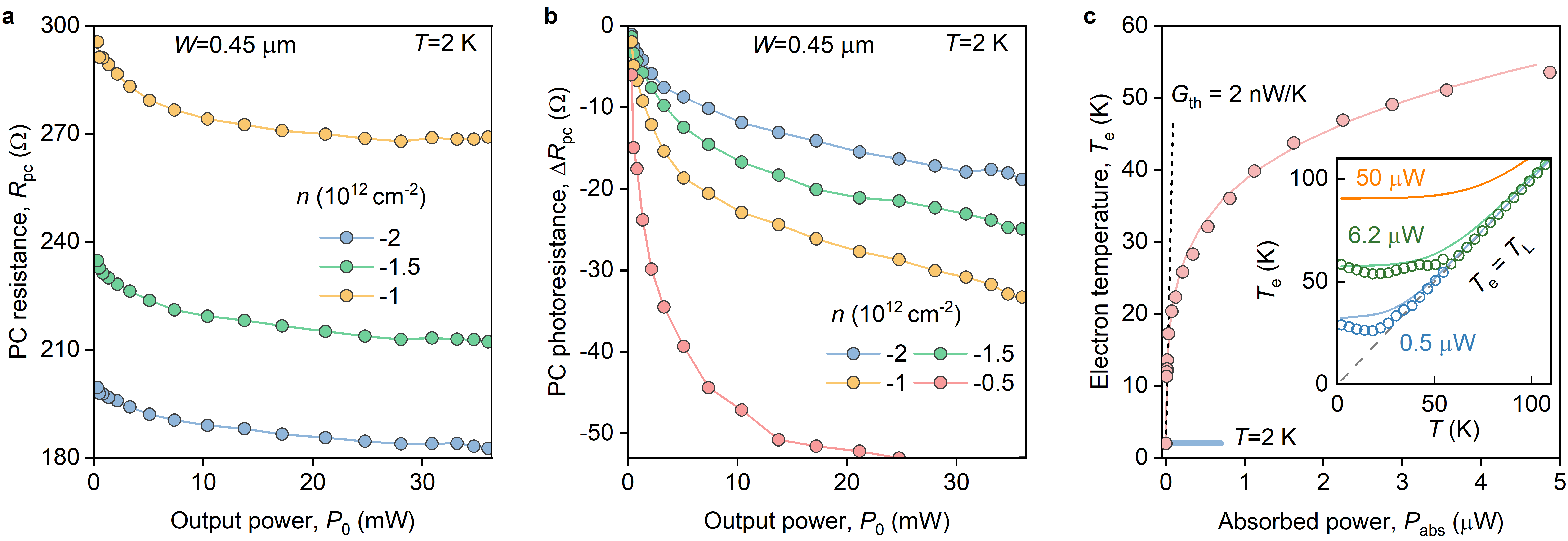}
    \caption{\textbf{Power dependence and superballistic electron thermometry.} \textbf{a,} Examples of the $R_\mathrm{pc}$ vs $P_\mathrm{0}$ dependencies measured at $T=2~$K for a $W=0.45~\mu$m device at different $n$. \textbf{b,} Corresponding $\Delta R_\mathrm{pc}$ vs $P_\mathrm{0}$ traces at given $n$. Data for $W=0.25~\mu$m is shown in Supplementary Information. \textbf{c,} Electron temperature $T_\mathrm{e}$ as a function of absorbed radiation power $P_\mathrm{abs}$ for $n = 2.5 \times 10^{12} \mathrm{ cm^{-2}}$ obtained for $W=0.25~\mu$m PC. Solid line: best fit to the solution of the heat balance equation. Fitting parameters: $\delta=4.65$ and $\Sigma=1.6$~mW/K$^\delta$m$^2$ Dashed line: linear fit yielding low-temperature $G_\mathrm{th}=2~$nW/K. Inset: Electron temperature $T_\mathrm{e}$ vs lattice temperature $T$ obtained for $W=0.25~\mu$m device. Solid lines: solution of the heat balance equation eq.~(3). } 
    \label{fig:F3}
\end{figure*}

We have also studied the dependence of $R_\mathrm{pc}$ on the excitation power. To this end, we used a waveguide attenuator connected to the output of our sub-THz source. In Fig.~\ref{fig:F3}a,  we plot $R_\mathrm{pc}$ against the output power $P_0$ of our 0.14~THz generator measured in one of our PC samples. For the $W=0.45~\mu$m PC, we found that $R_\mathrm{pc}(P_0)$ dependencies are non-linear across the whole $n-$range accessible in our devices and follow approximately the $|\Delta R_\mathrm{pc}| \sim  P^\alpha$ power law with $\alpha\approx1/4$ slightly varying with $n$. For a narrower PC of $W=0.25~\mu$m in width, we observed a similar dependence when graphene is heavily doped above $n=2\times 10^{12}~$ cm$^{-1}$ (Supplementary information). At smaller $n$, the $\Delta R_\mathrm{pc}$ dependence becomes somewhat non-monotonic: negative $\Delta R_\mathrm{pc}$ first decreases, reaches local minima, then starts to grow and changes sign at high excitation powers (Supplementary Information). 

\textbf{Superballistic electron thermometry.} The negative superballistic photoresistance presented in Figs.~\ref{fig:F2}a-c allows one to measure the increase of $T_\mathrm{e}$ upon THz absorption. To this end, we notice that the measured $\Delta R_\mathrm{pc}$ data depend on $T_\mathrm{e}$ through the following expression:
\begin{equation}\label{anti-M}
\Delta R_\mathrm{pc} (T)=\frac{1}{G_0+G_\mathrm{\nu}(\nu(T_\mathrm{e}))}-\frac{1}{G_0}.
\end{equation}
We then used experimentally determined values of $\nu(T_e)$, reported in ref.~\cite{kumar2017superball}, to find a correspondence between $\Delta R_\mathrm{pc}$ and $T_\mathrm{e}$. For both devices, we found that absorbed 0.14~THz radiation can elevate $T_\mathrm{e}$ by at least $\sim30-50~$K above the cold lattice (see also Fig.~2b, orange line) stimulating hydrodynamic electron flow free from momentum-relaxing e-ph collisions that are otherwise unavoidable in transport experiments where electrons are always thermalized with the lattice~\cite{bandurin2016,bandurin2018}. 

Next, it is instructive to determine the absorbed THz power, $P_\mathrm{abs}$. To this end, we notice that one can use Joule heating provided by the external dc current $I_\mathrm{dc}$ to achieve hydrodynamic electron flow~\cite{de_jong_hydrodynamic_1995}. By relating the THz-driven $R_\mathrm{pc}$ drop to the current-induced decrease of the differential resistances, $dV_\mathrm{pc}/dI$, measured as a function of  $I_\mathrm{dc}$ in the dark, we obtained the power calibration curve and revealed that only a small fraction of the total incident power is absorbed by our antenna-coupled devices (Supplementary Fig.X). 
This analysis allowed us to find the $T_\mathrm{e}$ dependence on $P_\mathrm{abs}$ presented in Fig.~3c for the $W=0.25~\mu$m PC. The data for another PC is shown in Supplementary Information. The obtained $T_\mathrm{e}(P_\mathrm{abs})$ dependencies experience a steep rise in the vicinity of zero $P_\mathrm{abs}$. Following already a standard analysis~\cite{Crossno,KC_noise,Specific_heat_Efetov}, we fitted the experimental data with the linear $P_\mathrm{abs}=G_\mathrm{th}(T_\mathrm{e}-T)$ dependence, where $G_\mathrm{th}$ is the thermal conductance describing the total heat escape from graphene to the bath, and obtained $G_\mathrm{th}\approx2~$nW/K (dashed line in Fig.~3c). This value agrees with previous experiments conducted using Johnson noise thermometry in the linear regime~\cite{KC_noise,Specific_heat_Efetov}. 

Upon further increasing $P_\mathrm{abs}$, the growth of $T_\mathrm{e}$ slows down with some tendency to saturate when $P_\mathrm{abs}$ exceedes $1~\mu$W. The observed non-linear $T_\mathrm{e}(P_\mathrm{abs})$ dependencies indicate that at $T_\mathrm{e}\gg T$, the hydrodynamic electron fluid driven out of equilibrium with the lattice relaxes heat via acoustic phonon emission~\cite{hot_carriers}. This can be demonstrated following the analysis from ref.~\cite{Specific_heat_Efetov}, namely via fitting the
$T_\mathrm{e}(P_\mathrm{abs})$ dependence by the solution of the heat balance
equation (See Supplementary Information for details):
\begin{equation}\label{heat}
P_\mathrm{abs}/A=-\nabla (\kappa \nabla T_\mathrm{e}) + \Sigma \left( T_e^\delta - T^\delta \right).
\end{equation}
Here, $\kappa$ is the in-plane thermal conductivity of graphene, $A$ is the device area, $\Sigma$ is the e-ph coupling constant. Using $\kappa$ from the Wiedeman Franz law we fit our data using the following fitting parameters: $\Sigma=1.6$~mW/K$^\delta$m$^2$ and $\delta=4.65$ (solid line in Fig.~3c) that are close to the values found in the Johnson thermometry experiments on graphene~\cite{Betz2013,Specific_heat_Efetov}.



It is also instructive to analyze electron overheating as a function of $T$ for various $P_\mathrm{abs}$. To this end, we notice that the measured $\Delta R_\mathrm{pc}$ data depend on $T$ only through the dependence of $T_\mathrm{e}$ on $T$. Indeed, momentum-non-conserving $T-$dependent e-ph scattering, which inevitably activates upon increasing $T$, contributes equally to $R_\mathrm{pc}$ measured in the dark and upon THz exposure and, therefore, does not enter $\Delta R_\mathrm{pc}$. This allows one to follow the analysis introduced above and by using experimental data from Figs.~2 and 3 together with eq.~(2) to plot $T_\mathrm{e}$ versus $T$ (Fig. 3c inset) that finds good agreement with the e-ph cooling model from eq.~(3) (solid lines in the inset of Fig. 3c). 

\textbf{Conclusion and outlook.} In conclusion, we observed anomalous (negative) photoresistance in metallic graphene at terahertz frequencies. This deviation from conventional behavior was caused by the THz-activated superballistic flow of hydrodynamic electrons through graphene PCs. We characterized the negative photoresistance as a function of carrier density, radiation power, and lattice temperature and demonstrated that superballistic transport can be used as an electron thermometer for THz-driven experiments on graphene. Using this approach, we demonstrated that absorbed radiation can thermally decouple electrons from the cold lattice by 50 K which is limited by the acoustic-phonon cooling. 

Our work also carries several important implications for the future design of THz-sensitive graphene-based devices. First, it provides a protocol for the conversion of absorbed THz power into the resistance change in graphene, illustrating the first practical use case of electron hydrodynamics. Second, THz-driven hydrodynamic electron transport provides a straightforward approach for measuring the amount of THz power, absorbed by antenna-coupled graphene devices, which remained unattainable in previous THz photocurrent experiments. Our approach highlights that, in typical experiments on antenna-coupled graphene and after accounting for losses in the optical path, the detector absorbs as little as $0.1~\%$ of impinging power. The primary reason is the impedance mismatch between the broadband antenna and the graphene channel: the characteristic radiation resistance of typical bow-tie antennae is only 50-100~$\Omega$ at 0.14~THz while the two-terminal resistance of graphene device is 5-10~k$\Omega$. As a result, almost all incident power is reflected from the device. Third, our $T_\mathrm{e}$ vs $P_\mathrm{abs}$ measurements highlight the importance of phonon cooling in THz graphene devices exploiting radiation to electronic heat conversion. Since such cooling, occurs at the picosecond scale, hydrodynamic electron bolometers can thus operate as ultra-fast THz sensors and mixers.

\vspace{1em}

\noindent\rule{6cm}{0.4pt}

*Correspondence to: dab@nus.edu.sg


 \section*{Data availability}
All data supporting this study and its findings are available within the article and its Supplementary Information or from the corresponding authors upon reasonable request.


\section*{Methods}

\textbf{Sample fabrication.} Our samples were fabricated from single-layer graphene films encapsulated by two hBN slabs using a standard dry-transfer technique described elsewhere\cite{CleaningInterfaces}. Narrow graphite strips were attached to the bottom surfaces of the stacks to serve as back gates. The samples were released onto the undoped insulating Si/SiO$_2$ substrates to ensure minimal THz power reflection. The Hall bar geometry of the devices, constriction in the middle of the channels and Ti/Au 1D metal contacts to graphene were defined using electron beam lithography followed by reactive ion etching in CHF$_3$/O$_2$ plasma. The source and drain electrodes were connected to the broadband triangular antennas capturing and directing the incident THz radiation into the samples' channels similar to previous studies~\cite{Bandurin_dual,bandurin2018}.  

\textbf{THz photoresistance measurements.} 
In our experiments, the samples were mounted in the 7 T variable temperature (1.8-300~K) optical cryostat manufactured by Quantum Design (Opticool). The 0.14 THz radiation (generated by Terasense source), was guided to the samples through the optical setup comprising lenses and a mirror. The waveguide attenuator installed at the output of the source allowed for a regulation of the THz powers reaching the samples.

To simultaneously measure the resistance and photoresistance of our devices, we applied an AC current of $I\sim$ 0.1 -- 2~$\mu$A at $f_\mathrm{I} = 33$~Hz, while modulating the THz source at $f_\mathrm{mod} = 82.5$~Hz. We used two lock-in amplifiers: one of them was phase-locked with the current source at $f_\mathrm{I}$, as in typical transport experiments, whereas the second operated in a dual modulation mode, measuring $V_\mathrm{dm}$ at $f_\mathrm{mod} - f_\mathrm{I}$ in a four-terminal configuration. The photoresistance is obtained through $\Delta R_\mathrm{pc} = \pi V_\mathrm{dm}/I$ representing the resistance change due to THz absorption. In this expression, an additional $\pi$ factor appears from the Fourier expansion of the rectangular signal modulating the source. A more detailed description of the setup and the double modulation technique is provided in the previous study~\cite{Artur_ACS}.\\

\section*{Competing interests}
The authors declare no competing interests.

\bibliography{Bibliography.bib}

\newpage
\setcounter{figure}{0}
\renewcommand{\thesection}{}
\renewcommand{\thesubsection}{S\arabic{subsection}}
\renewcommand{\theequation} {S\arabic{equation}}
\renewcommand{\thefigure} {S\arabic{figure}}
\renewcommand{\thetable} {S\arabic{table}}

\end{document}